\documentclass[UTF8,a4paper]{article}
\usepackage{amsmath}
\usepackage{graphicx}
\usepackage{subfigure}
\usepackage{epstopdf}
\usepackage{geometry}
\usepackage{ulem}
\usepackage{indentfirst}
\usepackage{amsfonts,amssymb,dsfont}
\usepackage{setspace}
\usepackage{threeparttable}
\usepackage{amsmath,mathtools,amsthm}
\usepackage{array}
\usepackage{extarrows}
\usepackage{upgreek}

\makeatletter
\renewcommand*\env@matrix[1][\arraystretch]{%
  \edef\arraystretch{#1}%
  \hskip -\arraycolsep
  \let\@ifnextchar\new@ifnextchar
  \array{*\c@MaxMatrixCols c}}
\makeatother
\begin{document}

%QED 3+1: Appelquist T W, Bowick M, Karabali D and Wijewardhana L C R 1986 Phys. Rev. D 33 3704
%Optical evidence for a Weyl semimetal state in pyrochlore Eu 2 Ir 2 O 7

\title{\bf Dynamical polarization and the optical response of silicene and related materials}
\author{Chen-Huan Wu
\thanks{chenhuanwu1@gmail.com}
\\Key Laboratory of Atomic $\&$ Molecular Physics and Functional Materials of Gansu Province,
\\College of Physics and Electronic Engineering, Northwest Normal University, Lanzhou 730070, China}

\maketitle
\vspace{-30pt}
\begin{abstract}
\begin{large} 

We discuss the dynamical polarization, optical response in low-frequency regime under in-plane polarized driving field
of the silicene.
The dynamical polarization, dielectric function, and absorption of radiation in infrared region are obtained
and shown in the ${\bf q}\sim\omega$ space,
and they are distinguishing for the cases of chemical potential larger than the band gap and smaller than the band gap.
The optical properties of silicene and the related group-V and group-VI materials: MoS$_{2}$ and black phosphorus are explored through the
first-principle study.
The plasmon which damped into the electron-hole pair in the single-particle excitation regime is also mentioned.
The spin/valley polarized electron-hole pairs can be formed through that way,
especially for the high-energy $\pi$-plasmon which begin to damp at the small ${\bf q}$-limit.
The
anisotropic effects induced by the warping structure or charged impurity, and the anisotropic 
polarization induced by the polarized incident light are also discussed.
Our result exhibits the great potential in the optoelectronic applications of the materials we discussed.\\

{\bf Keywords}: dynamical polarization; silicene; optical response; MoS$_{2}$; Black phosphorus; optical absorption\\

%{\bf PACS Nos.}: 31.10.+z; 31.15.Gy; 31.30.Jv\\

\end{large}

\end{abstract}
\begin{large}
\section{Introduction}

The optical response in periodically driven topological systems
is important for the study of the gain and loss mechanism, especially in the many-electron system.
For non-Hermitian system under laser, the conductivity and absorption as well as the scattering are related to the polarization vector
of the driving field even in the ${\bf q}\rightarrow 0$ limit,
where ${\bf q}$ is the in-plane Bloch wave vector in the direction of the in-plane light-polarization,
%{Linear optical properties in the projector-augmented wave methodology}
while in the imputity scattering system (with considerable Coulomb effect),
${\bf q}$ is the scattering wave vector (due to the charged impurities) which can be seen in the transition matrix element
of the dynamical polarization.
We note that the direction of the polarization is importnt for the optical or scattering properties of the two-dimension system
we talking in this article,
the anisotropic polarization induced by both the warping structure and the polarized incident light is discussed.
The hexagonal Dirac-cone warping leads to the anisotropic decay of the quasiparticle interference and the backscattering,
and it help to preserve the quasiparticle chirality.
%{Observation of Dirac cone warping and chirality effects in silicene}
While the anisotropic polarized light also leads to anisotropic conductivity or mobility\cite{Liu H,Liu Y}.
In this article, we mainly discuss the silicene and the related group-V and group-VI materils with the in-plane polarized of the driving field,

\section{Model of silicene in optical system}

For the low-energy Dirac-Hamilatonian of silicene in tight-binding model,
it reads\cite{Wu C HX,Wu C H3,Wu C H1,Wu C H4,Wu C H7}
\begin{equation} 
\begin{aligned}
H(t)=&\hbar v_{F}(\eta\tau_{x}P_{x}(t)+\tau_{y}P_{y}(t))+\eta\lambda_{{\rm SOC}}\tau_{z}\sigma_{z}+a\lambda_{R_{2}}\eta\tau_{z}(P_{y}\sigma_{x}-P_{x}\sigma_{y})\\
&-\frac{\overline{\Delta}}{2}E_{\perp}\tau_{z}+\frac{\lambda_{R_{1}}}{2}(\eta\sigma_{y}\tau_{x}-\sigma_{x}\tau_{y})+M_{s}s_{z}+M_{c}
-\eta\tau_{z}\hbar v_{F}^{2}\frac{\mathcal{A}}{\Omega}+\mu,
\end{aligned}
\end{equation}
where 
$P_{x}(t)=k_{x}-\frac{e}{c}A_{x}(t)=k_{x}-\frac{e}{c}A{\rm sin}\Omega t$ with $A$ the scalar potential.
$E_{\perp}$ is the perpendicularly applied electric field, 
$a=3.86$ is the lattice constant,
$\mu$ is the chemical potential,
$\overline{\Delta}$ is the buckled distance between the upper sublattice and lower sublattice,
$\sigma_{z}$ and $\tau_{z}$ are the spin and sublattice (pseudospin) degrees of freedom, respectively.
$\eta=\pm 1$ for K and K' valley, respectively.
$M_{s}$ is the spin-dependent exchange field and $M_{c}$ is the charge-dependent exchange field.
$\lambda_{SOC}=3.9$ meV is the strength of intrinsic spin-orbit coupling (SOC) and $\lambda_{R_{2}}=0.7$ meV is the intrinsic Rashba coupling
which is a next-nearest-neightbor (NNN) hopping term and breaks the lattice inversion symmetry.
$\lambda_{R_{1}}$ is the electric field-induced nearest-neighbor (NN) Rashba coupling which has been found that linear with the applied electric field
in our previous works\cite{Wu C H1,Wu C H2,Wu C H3,Wu C H4,Wu C H5}, which as $\lambda_{R_{1}}=0.012E_{\perp}$.
We here ignore the effects of the high-energy bands on the low-energy bands.
%{Photoinduced quantum spin and valley Hall effects, and orbital magnetization in monolayer MoS2}
The Dirac-mass and the corresponding quasienergy spectrum (obtained throught the diagonalization procedure) are\cite{Wu C HX}
%{Photoinduced quantum spin and valley Hall effects, and orbital magnetization in monolayer MoS2}
%{Photoinduced pseudospin effects in silicene beyond the off-resonant condition}
%{Spin valleytronics in silicene_ Quantum spin Hall–quantum anomalous Hall insulators and single-valley semimetals}
%{Valley-Polarized Metals and Quantum Anomalous Hall Effect in Silicene}
%{Photoinduced Topological Phase Transition and a Single Dirac-Cone State in Silicene}
%{Superconducting proximity effect in silicene- Spin-valley-polarized Andreev reflection, nonlocal transport, and supercurrent}
\begin{equation} 
\begin{aligned}
&m_{D}^{\eta\sigma_{z}\tau_{z}}=|\eta\sqrt{\lambda_{{\rm SOC}}^{2}+a^{2}\lambda^{2}_{R_{2}}k^{2}}s_{z}\tau_{z}-\frac{\overline{\Delta}}{2}E_{\perp}\tau_{z}+M_{s}s_{z}-\eta\hbar v_{F}^{2}\frac{\mathcal{A}}{\Omega}|,\\
&\varepsilon=s\sqrt{a^{2}\lambda^{2}_{R_{2}}k^{2}+(\sqrt{\hbar^{2}v_{F}^{2}{\bf k}^{2}
+(\eta\lambda_{{\rm SOC}}s_{z}\tau_{z}-\frac{\overline{\Delta}}{2}E_{\perp}\tau_{z}-\eta\hbar v_{F}^{2}\frac{\mathcal{A}}{\Omega} )^{2}}+M_{s}s_{z}+s\mu)^{2}},
\end{aligned}
\end{equation}
respectively, where the dimensionless intensity $\mathcal{A}=eAa/\hbar$ is the vector potential 
%in a form similar to the Bloch frequency, 
and $s=\pm 1$ is the electron/hole index.

The optical response in periodically driven topological systems (time-Floquet systems) which is non-Hermiticity, has been widely studied\cite{Koutserimpas T T}
where the existence of $\mathcal{PT}$-symmetry relys on the frequency of applied light.
It's believed that the related events through a certain operator $\hat{f}$ has the form of $B=\hat{f}A$\cite{Li C C} in quantum system,
e.g., through the SU(2) evolution operator, the effective Floquet Hamiltonian of the time-Floquet systems consider the Berry connection (Berry gauge potential)
reads, 
\begin{equation}
\begin{aligned}
H_{{\rm eff}}=\frac{i\hbar}{T}{\rm log}[\hat{\mathcal{T}}e^{-\frac{i}{\hbar}\int^{T}_{0}H(t)dt}]
-\frac{i\hbar}{T}{\rm log}[\hat{\mathcal{P}}e^{\frac{i}{\hbar}\int_{\mathcal{C}}A({\bf k})d{\bf k}}],
\end{aligned}
\end{equation}
where $\hat{\mathcal{T}}$ is the time-order operator $\hat{\mathcal{P}}$ is the path operator of the electron contour $C$
in the phase-space.
Note that here the contour encircled by the quasiparticle in momentum space and with the time-dependence of the Bloch bands
which is required by the anomalous velocity term due to the Berry curvature. 
For the case of light-driven contour in momentum space, $\mathcal{C}_{k}=\eta N\hbar\Omega$,
where $\mathcal{C}_{k}$ is the projection of $\mathcal{C}$ on the momentum space,
and $N$ is the number of Fourier modes.
%{Photoinduced Pseudospin Dynamical Effects in Graphene-Like Systems}
%{Photoinduced pseudospin effects in silicene beyond the off-resonant condition}
%{Photoinduced quantum spin and valley Hall effects, and orbital magnetization in monolayer MoS2}
The eigenfunction of system satisfies $\Psi(t)=e^{iN\hbar\Omega t}\psi(t)$,
and with the eigenvalue in the diagonal block of Floquet Hamiltonian shifted by the value of $\eta N\hbar\Omega$,
%{Hierarchy of Floquet gaps and edge states for driven honeycomb lattices}
while the time-dependent monochromatic harmonic perturbation enters in 
the off-diagonal blocks of the Floquet Hamiltonian\cite{Wu C H2,Wu C H5,Perez-Piskunow P M}.
The total Hamiltonian consider the effect of circularly polarized light is $H=\frac{1}{T}\int^{T}_{0}H(t)e^{-iN\hbar\Omega t}dt+V(t)$,
where $V(t)$ is the electron-radiation interaction
\begin{equation}
\begin{aligned}
V(t)=e^{-i\eta N\hbar\Omega t}(-ie{\bf v}A)+e^{i\eta N\hbar\Omega t}(-ie{\bf v}A)^{\dag},
\end{aligned}
\end{equation}
%{Universal absorption of two-dimensional materials within k · p method}
%{Photoinduced Pseudospin Dynamical Effects in Graphene-Like Systems}
where $A=E/\hbar\Omega$ is the scalar potential with $E$ the complex amplitude of electric field,
and ${\bf v}$ is the interband transition velocity matrix element,
which reads\cite{Wu C HX}
%{Spin-valley optical selection rule and strong circular dichroism in silicene}
%{Coupled Spin and Valley Physics in Monolayers of MoS2 and Other Group-VI Dichalcogenides}
\begin{equation} 
\begin{aligned}
{\bf v}=&\langle \psi|\frac{\partial H}{\hbar\partial {\bf k}}|\psi' \rangle +\hbar\partial_{t}{\bf k}\cdot\Omega({\bf k})\\
=& (v_{F}+\frac{i\eta a\lambda_{R_{2}}}{\hbar})(1+\eta\frac{m_{D}^{\eta s_{z}\tau_{z}}}{\varepsilon})\\
&+[\partial_{r}V({\bf r})+\frac{e}{\hbar}\partial_{\mu}\Phi({\bf k})e^{\mu}-\frac{e}{\hbar}{\bf v}_{g}\times{\bf B}({\bf k})]
\cdot\frac{1}{2 }\frac{2\eta \hbar^{2}v_{F}^{2}m^{\eta s_{z}\tau_{z}}_{D}}{2\varepsilon(4(m^{\eta s_{z}\tau_{z}}_{D})^{2}+\hbar^{2}v_{F}^{2}k^{2})},
\end{aligned}
\end{equation}
when takes the Berry correction into consideration,
where $\psi$ is the Bloch wave function in conduction band while $\psi'$ is the Bloch wave function in conduction band,
and $\Omega({\bf k})$ is the Berry curvature which is nonzero when the inversion symmetry is broken,
it reads
\begin{equation}
\begin{aligned}
\Omega({\bf k})=-{\rm Im}\left[
\sum_{\psi'\neq\psi}\frac{\langle\psi'|\partial_{k}H_{k}|\psi\rangle\times\langle\psi'|\partial_{k}H_{k}|\psi\rangle}{(\varepsilon_{\psi}-\varepsilon_{\psi'})^{2}}\right].
\end{aligned}
\end{equation}
Although the Berry curvature has been ignored in most of the computation of the velocity operator,
however, it's very important to the non-adiabatic correction in the system with broken reversal symmetry or the time-reversal symmetry 
(due to the off-resonance light
or the competition between Zeeman coupling and Rashba-coupling),
and it's closely related to the quantum anomalous Hall effect\cite{Zhao A,Jungwirth T},
%{Mapping the Berry curvature from semiclassical dynamics in optical lattices}
especially for the nonrelativity particle in semiclassical limit.
%{Enlarged Galilean symmetry of anyons and the Hall effect}
In the absent of Berry correction, it's been found that $\partial H_{cv}/\hbar\partial k=2\partial H_{cc}/\hbar\partial k$,
where $H_{cv}$ describes the interband transition and $H_{cc}$ describes the intraband transition.
Considering the Berry correction,
for the velocity of intraband transition, the berry term vanishes according to above expression (since $\psi=\psi'$),
thus $\partial H_{cv}/\hbar\partial k\neq 2\partial H_{cc}/\hbar\partial k$.

\section{Optical absorption in the presence of Dirac-cone warping}

The unpolarized optical absorption coefficient for spin- and valley-degenerate reads\cite{Huang R}
\begin{equation} 
\begin{aligned}
\alpha(\Omega)=\frac{g_{s}g_{v}}{2n_{t}}\alpha_{0}\int_{\mathcal{C}=\hbar\Omega}d\phi=\frac{4\pi}{n_{t}}\alpha_{0},
\end{aligned}
\end{equation}
where $\alpha_{0}=\frac{e^{2}}{2\epsilon_{0}h c}=1/137.036$ is the Sommerfeld fine structure constant,
%{Nair R R, Blake P, Grigorenko A N, et al. Fine structure constant defines visual transparency of graphene[J]. Science, 2008, 320(5881): 1308-1308.}
$g_{s}g_{v}$ denotes the spin and valley degenerate, $n_{t}$ is the refractive index.
For the special case that at the Dirac-point (${\bf k}=0$) with gapless band structure
and vanishing reflectivity coefficient and vanishing SOC\cite{Matthes L},
since then the transition is dominated by the intraband (mainly the conduction band) transition,
the well known optical absorption coefficient can be obtained as
\begin{equation} 
\begin{aligned}
\alpha(0)=\frac{g_{s}g_{v}}{2n_{t}}\alpha_{0}\pi=\frac{\pi}{n_{t}}\alpha_{0}=\frac{e}{n_{t}4\hbar\varepsilon_{0}c},
\end{aligned}
\end{equation}
where $\varepsilon_{0}$ is the permittivity of vacuum and $c$ is the speed of light.
Note that this ideal optical absorption coefficient only correct for the case that without the interband transition and SOC,
i.e., both the frequency and chemical potential need to be zero (undoped),
and for the chiral fermions with the gapless band structure.
%The degenerate of spin and valley is absent here unlike the graphene, and they are taken into account in the energy $\varepsilon$.
%{Universal absorption of two-dimensional materials within k · p method}
While in the semiclassical limit, the above result becomes
\begin{equation} 
\begin{aligned}
\alpha^{*}(0)=\frac{g_{s}g_{v}}{2n_{t}}\alpha_{0}\int_{\mathcal{C}=\hbar\Omega}
[1+(\partial_{t}{\bf k}\times\Omega({\bf k}))^{2}\hbar^{2}(\partial_{{\bf k}}\varepsilon)^{-2}
+2\hbar(\partial_{{\bf k}}\varepsilon)^{-1}{\bf k}\times\Omega({\bf k})]d\phi.
\end{aligned}
\end{equation}
In fact, in the presence of symmetry breaking (like the inversion symmetry or time-reversal invariance required by the nonzero Berry curvature),
the gap is generally nonzero,
thus the above ideal absorption coefficient $\frac{\pi}{n_{t}}\alpha_{0}$ is hard to realized
since the effect of interband transition is dominate for the gapped case.
The optical absorption between the $\pi$ and $\pi^{*}$ band can also be probed by the high-harmonic spectroscopy method
or the synchrotron radiation source\cite{Kobayashi K}.
%{Probing the π-π? transitions in conjugated compounds with an infrared femtosecond laser}
It's also found that, in a small frequency, the two-dimension universal absorbance becomes\cite{Matthes L2}
\begin{equation} 
\begin{aligned}
\alpha(0)=\frac{g_{s}g_{v}}{n_{t}}[\alpha_{0}\pi(1+\frac{1}{16t^{2}}\hbar^{2}\omega^{2})],
\end{aligned}
\end{equation}
where $t$ is the nearest-neighbor hopping modified by the light, and becomes $1.09$ eV $\sim$ $0.4$ eV
($t=1.6$ eV for $\omega=0$ case).

In the low-energy-limit, we usually consider only the direct interband transition,
however, the indirect interband exist mediated by the remote band\cite{Huang R},
which gives rise the warping effect.
Unlike the graphene or MoS$_{2}$\cite{Kormányos A} which have trigonal warping (with three Dirac-cone) constant-energy-contour (or the Dirac-cone) 
at high energy state\cite{Tikhonenko F V},
the silicene on Ag(111)\cite{Feng B} or other substrates has hexagonl warping 
with six Dirac-cone for the Dirac-cone or local density of states (LDOS) both in momentum space and real space,
like the Bi$_{2}$Te$_{3}$\cite{Fu L}.
%{Observation of Dirac cone warping and chirality effects in silicene}
The plots are presented in the Fig.1,
where we can see that the hexagonal warping is more obvious in LDOS for large $t$.
We also can see that, the lower the energy, the higher the degree of isotropy of the constant-energy-contour in Brillouin zone (BZ),
and thus with higher isotropy of quasiparticle scattering.
%{Observation of Dirac cone warping and chirality effects in silicene}
The hexagonal warping produces strongly decays the quasiparticle interference\cite{Feng B} as well as
the backscattering,
thus it retard the decaying of the LDOS (the Friedel oscillation) and breaks the weak intravalley localization\cite{Tikhonenko F V}.
The quasiparticle interference here is due to the broken of quasiparticle chirality.
%{Observation of Dirac cone warping and chirality effects in silicene}
%{Weak Localization in Graphene Flakes}
The intervalley scattering which is suppressed in the clean limit corresponds to the short-wavelength interference (large ${\bf k}$)
and it's competing with the quasiparticle chirality,
while the intravalley scattering corresponds to the long-wavelength interference (small ${\bf k}$).
For time-reversal-invariant system, the opposite spin (or pseudospin)
also suppress the intervalley backscattering in a warping system.
The saddle-point Van Hove-singularities are shown in the LDOS plot of Fig.1 which are localed in the $M$-point of the BZ boundary,
and also generates the Fermi surface by the quasiparticle scattering along the Fermi patches between two opposite edges\cite{Feng B}.
The DOS is not
differentiable in these points due to the saddle point singularity\cite{John R}.
%{Optical properties of graphene, silicene, germanene, and stanene from IR to far UV–a first principles study}
However, for the case of large gap and in the presence of remote-band coupling\cite{Huang R},
the minimum-point Van Hove-singularities\cite{Souslov A} appear, which we don't discuss in this article.
%{Universal absorption of two-dimensional materials within k p method}
For bilayer silicene or bilayer graphene, another source of the trigonal Warping\cite{Ezawa M,McCann E,Morell E S}
is the interlayer hopping which has three direction for the hopping from one site in bottom layer to the upper layer,
%[Quasi-Topological Insulator and Trigonal Warping in Gated Bilayer Silicene}
and it's not negligible especially under the laser in terahertz range.
%{Radiation effects on the electronic properties of bilayer graphene}
%{McCann E, Fal’ko V I. Landau-level degeneracy and quantum Hall effect in a graphite bilayer[J]. Physical Review Letters, 2006, 96(8): 086805.}

\section{Dynamical polarization in low-frequency within random phase approximation}

It's found that the low-frequency (infrared or visible region) absorbance is similar among the two-dimension group-IV crystals
\cite{Matthes L2},
like the graphene, silicene, and germanene, we only take silicene as an example here.
Firstly, through the random phase approximation (RPA) in the presence of strong SOC,
the dielectric function within one-loop approximation reads
\begin{equation} 
\begin{aligned}
\varepsilon(\omega,{\bf q})=1-\frac{2\pi e^{2}}{\epsilon_{0}\epsilon{\bf q}}\Pi(\omega,{\bf q}),
\end{aligned}
\end{equation}
where $\epsilon=2.45$ is the static background dielectric constant for the air/SiO$_{2}$ substrate,
and $\Pi(\omega,{\bf q})$ is the dynamical polarization function\cite{Wu C H3}
\begin{equation} 
\begin{aligned}
\Pi(\omega,{\bf q})=g_{s}g_{v}\frac{2\pi e^{2}}{\epsilon_{0}\epsilon}\sum_{m_{D}}
\int_{BZ}\frac{d^{2}k}{4\pi^{2}}\sum_{{\bf q};s,s'=\pm 1}\frac{f_{s({\bf k}+{\bf q})}-f_{s'{\bf k}}}{s\varepsilon_{{\bf k}+{\bf q}}
-s'\varepsilon_{{\bf k}}-\omega-i\delta}{\bf F}_{ss'}({\bf k},({\bf k}+{\bf q})),
\end{aligned}
\end{equation}
%{Dynamic screening and low-energy collective modes in bilayer graphene}
%{Frequency-dependent polarizability, plasmons, and screening in the two-dimensional pseudospin-1 dice lattice}
%{Screening, Kohn Anomaly, Friedel Oscillation, and RKKY Interaction in Bilayer Graphene}
%{Universal infrared absorbance of two-dimensional honeycomb group-IV crystals}
where 
$s,s'$ are the band index ($ss'=1$ for the intraband case and $ss'=-1$ for the interband case),
and
the Coulomb interaction-induced transition matrix element here is
\begin{equation} 
\begin{aligned}
{\bf F}_{ss'}({\bf k},({\bf k}+{\bf q}))=\frac{1}{2}(1+ss'{\rm cos}\theta_{\sigma\eta})=\frac{1}{2}\left[1+ss'(\frac{{\bf k}({\bf k}+{\bf q})}{E_{{\bf k}}E_{{\bf k}+{\bf q}}}
+\frac{m_{D}^{2}}{E_{{\bf k}}E_{{\bf k}+{\bf q}}})\right],
\end{aligned}
\end{equation}
where $\theta_{\sigma\eta}$ is the scattering angle in scattering phase space 
%={\rm arctan}\frac{\eta\hbar v_{F}{\bf k}}{2m_{D}}
%{Mobility anisotropy in monolayer black phosphorus due to scattering by charged impurities}
%{Scattering in graphene associated with charged out-of-plane impurities}
where the anisotropic intervalley scattering is possible through the edge states.
%{Anisotropic Friedel oscillations in graphene-like materials The Dirac point approximation in wave-number dependent quantities revisited}
%Here the Dirac-mass is related to the band gap in Dirac-cone by $|2m_{D}|= \Delta$\cite{Wu C H2}.
Here the transition matrix element ${\bf F}_{ss'}({\bf k},({\bf k}+{\bf q}))$ including both the interband ($ss'=-1$) and intraband ($ss'=1$) transitions.
%While in the simply circular coordinate system.
%and it shows the pseudospinorial character of electrons\cite{Matthes L2}. 
The scattering angle in momentum space between ${\bf k}$ and ${\bf k}+{\bf q}$ is $\theta$ and obeys
${\rm cos}\theta=\langle\chi({\bf k})|\chi({\bf k}+{\bf q})\rangle=(k+q{\rm cos}\phi)/\sqrt{k^{2}+q^{2}+2kq{\rm cos}\phi}$ 
%{Dynamic screening and low-energy collective modes in bilayer graphene}
%{Frequency-dependent polarizability, plasmons, and screening in the two-dimensional pseudospin-1 dice lattice}
where $\phi$ is the angle between ${\bf k}$ and ${\bf q}$, and $|\chi({\bf k})\rangle=\psi_{s}^{*}({\bf k})\psi_{s'}({\bf k})$,
$|\chi({\bf k}+{\bf q})\rangle=\psi_{s}({\bf k}+{\bf q})\psi_{s'}^{*}({\bf k}+{\bf q})$ is the eigenstates with the eigenvectors $\psi$ of the Hamiltonian
\cite{Wu C H3}.
%Note that here the scalar product are the simplification of $\langle\chi({\bf k})|\int^{\pi/a}_{-\pi/a}e^{-i{\bf q}r{\rm cos}\theta}d\theta|\chi({\bf k}')\rangle=
%\langle\chi({\bf k})|\chi({\bf k}')\rangle \delta({\bf k}',{\bf k}+{\bf q})$.

\subsection{$\mu=2$ eV$>\Delta$}

Firstly, we discuss the case of chemical potential $\mu=2$ eV which is larger than the band gap.
In the ${\bf q}\sim\omega$ space formed by the scattering wave vector ${\bf q}$ and frequency $\omega$,
the single particle excitation regime (or electron-hole continuum regime) can be devided into two parts:
The intraband part and interband part
as shown in Fig.2,
where the blue line surrounds the interband part while the red line surrounds the low-energy intraband part.
The region surrounded by the blue line and the region surrounded by the red line can be analytically expressed as\cite{Wu C H3,Pyatkovskiy P K}
\begin{equation} 
\begin{aligned}
\mu+\sqrt{({\bf q}-{\bf k}_{F})^{2}+ (m_{D}^{\eta \sigma_{z}\tau_{z}})^{2}}<\omega<\mu+\sqrt{({\bf q}+{\bf k}_{F})^{2}+ (m_{D}^{\eta \sigma_{z}\tau_{z}})^{2}}\\
\omega<\mu-\sqrt{({\bf q}-{\bf k}_{F})^{2}+ (m_{D}^{\eta \sigma_{z}\tau_{z}})^{2}},
\end{aligned}
\end{equation}
respectively,
where ${\bf k}_{F}=\sqrt{\mu^{2}-(m_{D}^{\eta \sigma_{z}\tau_{z}})^{2}}$ is the Fermi wave vector.
Note that the above functions are valid for any value of chemical potential.

We can see that, for band gap $\Delta=2$ eV (i.e., equals to the chemical potential),
the spingle particle excitation regime vanishes in the region shown in the last panel of Fig.2 and thus the 
plasmon model won't undamped at all in this region.
%{Probing the topological phase transition via density oscillations in silicene and germanene}
The value of ${\bf q}=2{\bf k}_{F}$ decrease with the increasing band gap as indicated by the red arrow.
In Fig.3, we show the dynamical polarization at the ${\bf q}\sim\omega$ space.
%which corresponds to the red line in Fig.2 in the zero-frequency limit.
We want to note that the imaginary part of polarization function is not always negative as shown in the figure (see also the Refs.\cite{Wu C H3,Kotov V N}).
Fig.4 shows the dielectric function obtained above,
we can see the real part of dielectric function is much larger than the imaginary part,
and there is a peak in the real part along the line of ${\bf q}=2{\bf k}_{F}$ as indicated by the red arrow.
The Fig.5 shows the absorption of the radiation (not the optical absorbance),
which reads
\begin{equation} 
\begin{aligned}
\alpha({\bf q},\omega)=-{\rm Im}\frac{1}{\varepsilon({\bf q},\omega)}=-L({\bf q},\omega),
\end{aligned}
\end{equation}
where $L({\bf q},\omega)$ is the energy loss function.
%{Dynamical polarization function, plasmons, and screening in silicene and other buckled honeycomb lattices}
Thus it's easy to obtain that the energy loss function should be the inversion of the absorption,
and they are both dependent on the band gap and the spin-valley coupled selection rule\cite{Wu C HX}.
From Fig.5, the negative optical absorption appear for small band gap (e.g., $\Delta=0.02$ eV)
which shows no practical physical meaning.
Here we want to note that the energy loss function provides the electron-hole spectral
density in single-particle excitation regime, and the damping in single-particle excitation
regime also leads to the resonance of the energy loss function\cite{Wu C H3}.
The process of the energy-loss exist as long as the frequency $\omega$ is nonzero and it also results in the loss of DOS
just like the one caused by the chiral quasiparticle scattering though it's suppressed by the hexagonal warping.
%{Observation of Dirac cone warping and chirality effects in silicene}

The absorption vanishes in the last panel of Fig.5 as the single particle excitation regime vanishes (see Fig.2).
For the case of large band gap (in the adiabatical approximation),
the electron-ion plasmon rised in plasmon frequency and at the presence of long-range Coulomb interaction\cite{Wunsch B}.
In this case, the phonon dispersion of the acoustic phonon can be obtained as\cite{Wunsch B}
\begin{equation} 
\begin{aligned}
\omega_{{\rm ph}}=\sqrt{\frac{\alpha E_{ion}}{\epsilon_{0}\epsilon{\bf q}+g_{s}g_{v}\alpha\mu/4\pi}}{\bf q},
\end{aligned}
\end{equation}
where $E_{ion}$ is the ion confinement energy and $\alpha=e^{2}/\epsilon_{0}\epsilon\hbar v_{F}$ is the fine structure constant.
Through the expression, the phonon dispersion also follows the $\sqrt{{\bf q}}$-behavior as most two-dimension materials do.
%{Spin and valley polarization of plasmons in silicene due to external fields}
The phonon dispersion is shown in the dash-line of the first panel of Fig.5,
which mainly distributed in the low-frequency regime 
(while the optical phonon dispersion is mainly distributed in the high-frequency regime),
and independent of the band gap.

\subsection{$\mu=0.01$ eV$<\Delta$}

We then discuss the case that the chemical potential is smaller than the band gap, where we set it as 0.01 eV here.
In this case, the imaginary part of the dielectric function reads\cite{Tabert C J,Pyatkovskiy P K,Scholz A}
\begin{equation} 
\begin{aligned}
{\rm Im}[\varepsilon({\bf q},\omega)]=\frac{2\pi e^{2}}{\epsilon_{0}\epsilon{\bf q}}\frac{{\rm q}^{2}}{16}\theta(\omega^{2}-{\bf q}^{2}
-4(m_{D}^{\eta\sigma_{z}\tau_{z}})^{2})(\frac{1}{\sqrt{\omega^{2}-{\bf q}^{2}}}+\frac{4(m_{D}^{\eta\sigma_{z}\tau_{z}})^{2}}{(\omega^{2}-{\bf q}^{2})^{3/2}}).
\end{aligned}
\end{equation}
Through the Kramers-kronig relation
\begin{equation} 
\begin{aligned}
{\rm Re}[\Pi({\bf q},\Omega)]=\frac{2}{\pi}\mathcal{P}\int^{\infty}_{0}d\omega\frac{\omega{\rm Im}[\Pi({\bf q},\omega)]}{\omega^{2}-\Omega^{2}},
\end{aligned}
\end{equation}
the real part of the dielectric function can be obtained as
\begin{equation} 
\begin{aligned}
{\rm Re}[\varepsilon({\bf q},\omega)]=&\frac{2\pi e^{2}}{\epsilon_{0}\epsilon{\bf q}}\frac{{\rm q}^{2}}{4\pi}
[\frac{m_{D}^{\eta\sigma_{z}\tau_{z}}}{{\bf q}^{2}-\omega^{2}}+\frac{{\bf q}^{2}-\omega^{2}-4(m_{D}^{\eta\sigma_{z}\tau_{z}})^{2}}{4|{\bf q}^{2}-\omega^{2}|^{3/2}}\\
&(\theta({\bf q}-\omega){\rm arccos}\frac{{\bf q}^{2}-\omega^{2}-4(m_{D}^{\eta\sigma_{z}\tau_{z}})^{2}}{\omega^{2}-{\bf q}^{2}-4(m_{D}^{\eta\sigma_{z}\tau_{z}})^{2}}
-\theta(\omega-{\bf q}){\rm ln}\frac{(2m_{D}^{\eta\sigma_{z}\tau_{z}}+\sqrt{\omega^{2}-{\bf q}^{2}})^{2}}{|\omega^{2}-{\bf q}^{2}-4(m_{D}^{\eta\sigma_{z}\tau_{z}})^{2}|})].
\end{aligned}
\end{equation}
where $\theta$ is the step function.
In this case, the dynamical polarization, dielectric function, and the absorption are shown in the Fig.6-8, respectively.
We then obtain the purely negative dynamical polarization and purely positive dielectric function as shown in Fig.6-7.
Distincted from Fig.5, the absorption for $\mu<\Delta$ is purely positive, 
and mainly locate in the interband single particle excitation regime as shown in the Fig.8.
%{Spin and valley polarization of plasmons in silicene due to external fields}
That reveals that it's quite important for the value of chemical potential compared to the band gap
(and it's also important for the estimation of the effect of Berry correction as presented in the Sec.2).

\section{Optical properties of Silicene, MoS$_{2}$, and black phosphorus}

The scattering momentum here is mainly contributed from the charged impurity scattering (with Coulomb interaction),
while for the optical transition,
due to the almost vanishing photon wave vector, i.e., in the limit of ${\bf q}\rightarrow 0$,
which also called the head or wings of the polarizability\cite{Gajdo? M}.
We focus only on the interband optical transition through a finite gap, then the 
transition matrix becomes\cite{Matthes L2}
%{Optical properties of two-dimensional honeycomb crystals graphene, silicene, germanene, and tinene from first principles.pdf}
\begin{equation} 
\begin{aligned}
\sqrt{{\bf F}({\bf k})}
=\frac{e\hbar}{i\sqrt{4\pi\epsilon_{0}}m_{0}}\frac{\langle\psi;{\bf k}|m_{0}{\bf v}\cdot{\bf e}_{\bf q}|\psi';{\bf k}\rangle}{\varepsilon_{\psi}-\varepsilon_{\psi'}},
\end{aligned}
\end{equation}
where ${\bf v}$ is the velocity operator and ${\bf e}_{\bf q}$ denotes the direction of scattering wave vector which is
also the direction of the dielectric function.
At Dirac-cone (i.e., ${\bf k}=0$), the velocity operator in above can be written in the same form as Eq.(5) in the nonadiabatic approximation.

We carry out the density functional theory (DFT) calculation base on the generalized gradient approximation (GGA) in Quantum-ESPRESSO package\cite{Giannozzi P},
The plane wave energy
cutoff is setted as 400 eV for the ultrasoft pseudopotential in our calculations, and the structures
are relaxed until the Hellmann-Feynman force on the atoms are below 0.01 eV/\AA\ .
%{A first-principles investigation of the optical spectra of oxidized graphene}
The optical properties including the optical absorption has been studied for group IV matters, 
like the silicene, graphene, germanene and tinene\cite{Matthes L,Singh N},
%{Optical properties of graphene, silicene, germanene, and stanene from IR to far UV–a first principles study}
while in this section, we focus on several typical materials in group IV-VI: monolayer silicene, monolayer MoS$_{2}$, and monolayer black phosphorus (phosphorene),
%{Layer-controlled band gap and anisotropic excitons in few-layer black phosphorus}
which all have strong intrinsic SOC and hexagonal layered structure, and all exhibit abundant optical response characteristics\cite{Çakır D}.

The results of DFT calculation are shown in Fig.6,
where we show the in-plane component of the absorption, energy loss function, optical parameters, and the dielectric function of these three matter.
A clear splitting of the main peak in optical absorption is presented,
which can't be found in the energy loss function,
we think the reason for such difference is partly due to the local field effect (the off-diagonal element of the dielectric function),
which is been ignored in the approximation of ${\bf G}+{\bf q}\rightarrow 0$\cite{Gajdo? M} where ${\bf G}$ is the vectors of the reciprocal lattice.
%{Dynamical polarization of graphene under strain}
%{Pellegrino F M D, Angilella G G N, Pucci R. Dynamical polarization of graphene under strain[J]. Physical Review B, 2010, 82(11): 115434.}
The neglect of the local-field effect also results in the decrease of
the number of the plasmon branches due to the suppression of the intreband transition\cite{Wu C H3}.
For the system with quasi-one-dimensional band structure, like the black phosphorus\cite{Tran V},
the many-electron effect sometimes need to be taken into account except for in the low-temperature limit,
it's similar to the case as we discussed\cite{Wu C H2,Wu C H8}
with the considerable long- and short-range Coulomb effect.
Due to the many-electron effect, the band gap as well as the related excitonic effect are also affected by the self-energy matrix $\Sigma$\cite{Tran V},
a large nontrivial band gap may opened by the many-electron effect.
The variable self-energy also leads to nonzero vertex contribution
which is important especially in the variant cluster approximation.
%{Spatial Correlations and the Insulating Phase of the High- Cuprates Insights from a Configuration-Interaction-Based Solver for Dynamical Mean Field Theory} 
It's also found that, for superlattice arrangement of the layered structure, when ${\rm Re}\varepsilon({\bf q}\rightarrow 0,\omega)\approx 1$,
the shape of absorption and the energy loss function may be more similar\cite{Matthes L}.
A small broaden peak near 10 eV in absorption is observed both for silicene and black phosphorus,
it's due to the transition between the $s$ hybridized orbital and the $\pi^{*}$ band
in parallel band region for group-IV materials.
%{Optical properties of graphene, silicene, germanene, and stanene from IR to far UV –A first principles study}

We found that at ${\bf q}\rightarrow 0$ limit, the initial refractive index $n_{t}$ descrise from monolayer silicene to monolayer black phosphorus:
4.05 eV, 18 eV, 2.15 eV for monolayer silicene, monolayer MoS$_{2}$, and monolayer black phosphorus, respectively,
which means that, in this order, the coupling between $\pi$-band and $\sigma$-band decrease while the electron mobility increase\cite{John R}.
Except for monolayer black phosphorus, the maximum refractive indices are almost appear in the zero photon energy.

For dielectric function, we see that the real part of all these three matters have negative part,
which means the existence of the plasmon frequency,
since the plasmon frequency can be approximatly obtained by solving the ${\rm Re}[\varepsilon({\bf q},\omega_{p})]$ for the case of small damping.
%{Dielectric function, screening, and plasmons of graphene in the presence of spin-orbit interactions}
Thus through the plots of dielectric function we can obtain the plasmon frequency of monolayer silicene, monolayer MoS$_{2}$, and monolayer black phosphorus as
8 eV, 18 eV, and 11 eV, respectively.
%{Optical properties of graphene, silicene, germanene, and stanene from IR to far UV –A first principles study}
These results also agree with the plots of energy loss function,
whose peak also reveals the plasmon frequency,
and we can see that the peaks are locate in 8 eV, 17.5 eV, 11 eV for the monolayer silicene, monolayer MoS$_{2}$, and monolayer black phosphorus,
respectively,
which are very close to the above results in the plots of dielectric function.
%{Optical properties of graphene, silicene, germanene, and stanene from IR to far UV – A first principles study}
That's to say, both the peaks of energy loss function and the dip of the real part of (in-plane) dielectric function
are due to the coherent collective excitation models.
%{Excitation spectrum and high-energy plasmons in single-layer and multilayer graphene}
Note that the plasmon frequency doesn't exist in the graphene which has purely positive real dielectric function
as presented in Refs.\cite{John R},
and thus exhibits pure metallic bahavior with with the optical response mainly in ultraviolet regime.
We can also see that the imaginary part of the dielectric function is in a similar shape with the absorption,
which means that the positive imaginary part of the dielectric function corresponds to the abosrption gain.

From the plot of energy loss function of silicene,
there are also two small peaks in front of the main peak in 8 eV,
which are in 2 eV and 5 eV, and contributed by the $pi$-plasmon and $\sigma$-plasmon, respectively.
Note that here the $pi$-plasmon peak is rather weak (lower than 0.3 eV),
however, in doped case, it's possible to produce the high energy $pi$-plasmon (about 5$\sim$6 eV) due to the Van Hove singularities,
as observed in doped graphene\cite{Yuan S},
such high energy $pi$-plasmon correponds to the large collective excitation which will
decay the plasmon into electron-hole pair,
and it's unlike the acoustic plasmon mention above which follows the ${\bf q}$-behavior,
but follows the linear behavior with ${\bf q}$ just like the classical bilayer silicene as discussed in our previous report\cite{Wu C H3}.
The reason of the linear behavior for high energy $pi$-plasmon is due to the fast damping in small ${\bf q}$ region or even in the ${\bf q}\rightarrow 0$ limit
\cite{Yuan S} and thus it's hard to find the stable plasmons for such case.
%{Excitation spectrum and high-energy plasmons in single-layer and multilayer graphene}
Note that all the optical parameters obtained here are smaller than that in their host materials (bulk form).
We are mainly focu on the infrare and visible region of the photon energy,
however, in negative photon energy, some interst phenomenons are been found,
like the multi-photon resonance due to the transition of subbands\cite{Yin X}.

\section{Conclusion}

The optical response is of special interest for the intriguing materials of the group-IV, group-V, and group-VI.
We diacuss the optical properties of the silicene, MoS$_{2}$ and black phosphorus,
and it's important for their exciting potential applications.
While in semiclassical case with small band gap (i.e., the non-adiabatic) case, the Berry correction is also important for the interband or intraband transition 
matrix element.
Except that, the opposite Berry curvature and spin/orbital magnetic moment between neighbor valley 
also give rise to the topological spin/valley Hall effect.
%{Giant Spin-valley Polarization and Multiple Hall Effect in Functionalized Bi}
For silicene, the dynamical polarization, dielectric function, and the absorption of the radiation are discussed in the absence of many-electron effect.
The many-electron effects on optical response need to be considered in the presence of segments of quasi one-dimension band, like the black phosphorus,
due to the presence of many-electron effect, the self-energy and the related excitonic effect need to be taken into account\cite{Yang L}.
In the presence of Coulomb interaction (electron-electron interaction), it also leads to the dampling of
self-energy of the Dirac-quasiparticle (due to the electron-phonon coupling, electron-collision,
or the acoustic phonon scattering) in the low-energy limit and it follows the Kramers-Kronig
relation, then the energy becomes
$\varepsilon\rightarrow \varepsilon+\Sigma,
\ {\rm Re}\Sigma \sim g_{c}e^{1/g_{c}},\ {\rm Im}\Sigma \sim g^{2}_{c}e^{1/g_{c}}$\cite{González J,Khveshchenko D V},
with first term the linear dispersion term and the second term $\Sigma$ the nonlinear dispersion term due to the screening effect.
The
anisotropic effects induced by the hexagonal warping structure of silicene or the charged impurity, and the anisotropic 
polarization induced by the polarized incident light are also discussed.
Our results exhibit the great potential in the optoelectronic applications of the materials we discussed.

\end{large}
\renewcommand\refname{References}

\clearpage
%\section{Table}
%Table 1:Three kinds of distorted structure of silicene which are transferred into the insulating phase with unambiguous band gaps
%compare to the pristine one.
%The corresponding bond angles between e-bond and f-bond and the (average) bulkling distances $\overline{\Delta}$ are also shown.
%The corresponding schematic of this displacing (distorting) was shown in the Fig.4(e).
%\begin{table}[!hbp]
%\centering
%%\resizebox{\textwidth}{!}{
%%\begin{threeparttable}
%%\begin{spacing}{1.19}
%\begin{tabular}{cccccccccc}
%\hline
%\hline  
%structure                   &  a (\AA)   &b (\AA)&c (\AA)&d (\AA)&e (\AA)&f (\AA)&$\overline{\Delta}$ (\AA) &Bond angle&Band gap (eV)\\
%\hline    
%pristine &2.280 &2.280 &2.280 &2.280 &2.280  &2.280 &0.51&115.96$^\text{o}$& $5.45\times 10^{-7}$\\
%distorted 1    &  2.245    &2.245&2.263&2.263&2.246&2.247&0.47 &116.821$^\text{o}$&1.609\\
%distorted 2    &  2.245    &2.245&2.263&2.263&2.245&2.246 &0.45&117.119$^\text{o}$&1.610\\
%distorted 3    &  2.245    &2.245&2.264&2.263&2.245&2.245 &0.42&117.247$^\text{o}$&1.612\\
%\hline
%\hline     
%\end{tabular}
%%\end{spacing}
%%\end{threeparttable}}
%\end{table}

\clearpage
Fig.1
\begin{figure}[!ht]
   \centering
 \centering
   \begin{center}
     \includegraphics*[width=0.7\linewidth]{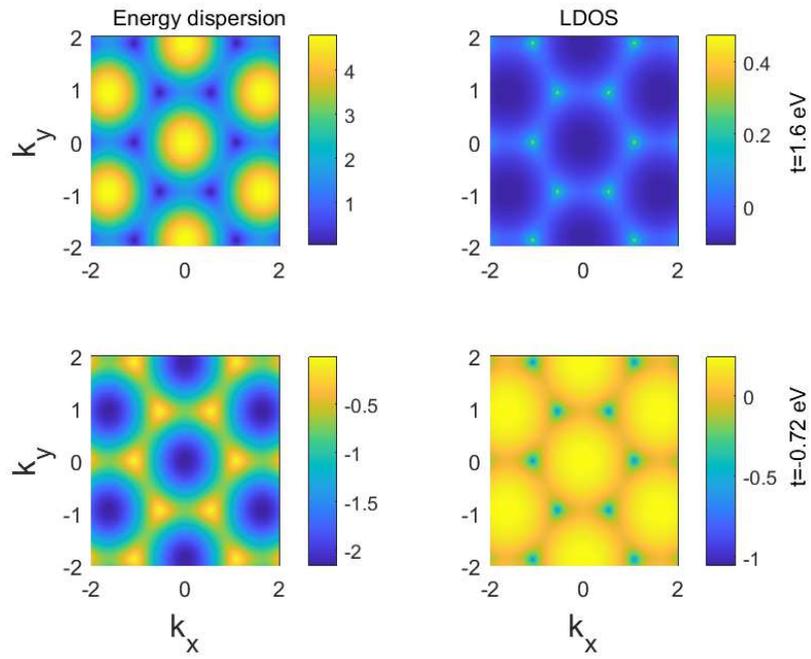}
\caption{(Color online) Constant-energy-contour (Fermi surface) of the tight-binding energy dispersion (left panels) and LDOS (right panels) for silicene.
The LDOS is obtained by the renormalization group method in momentum space.
The upper panels corresponds to $t=1.6$ eV, while lower panels corresponds to $t=-0.72$ eV 
(when only consider the nearest-neighbor $pp\pi$-band\cite{Guzmn-Verri G G}).
}
   \end{center}
\end{figure}
\clearpage

Fig.2
\begin{figure}[!ht]
   \centering
 \centering
   \begin{center}
     \includegraphics*[width=0.7\linewidth]{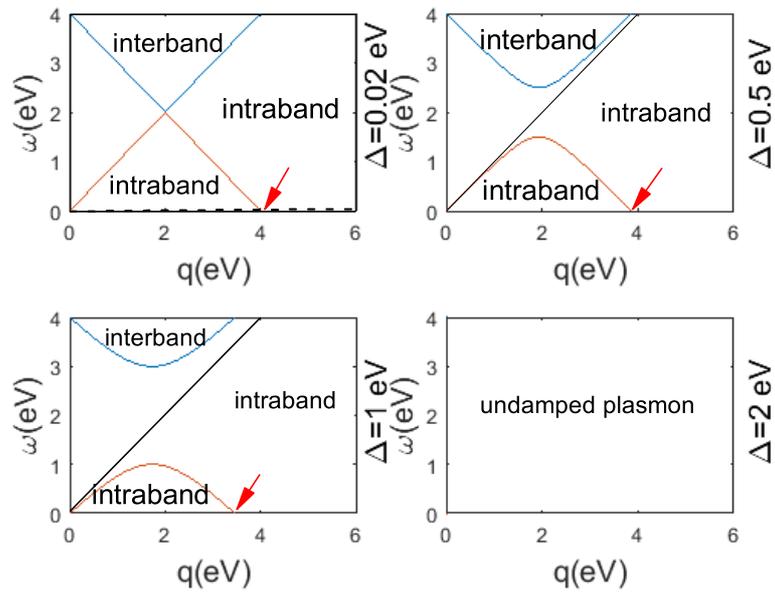}
\caption{(Color online) The intraband part and interband part of the single particle excitation regime in ${\bf q}\sim\omega$ space.
The red arrows indicate the point ${\bf q}=2{\bf k}_{F}$.
The dash-line in first panel is the  acoustic phonon dispersion with the electron-ion plasmon.
}
   \end{center}
\end{figure}
\clearpage
Fig.3
\begin{figure}[!ht]
   \centering
 \centering
   \begin{center}
     \includegraphics*[width=0.7\linewidth]{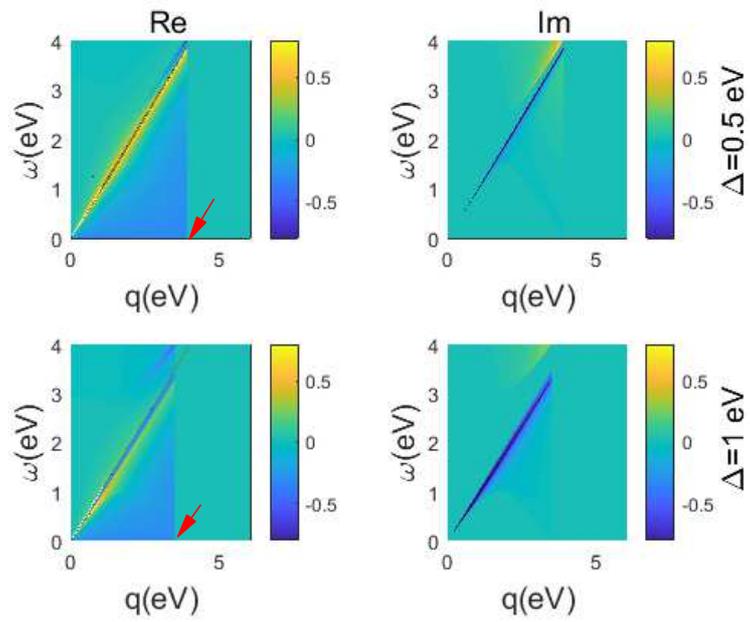}
\caption{(Color online) Dynamical polarization at finite scattering wave vector ${\bf q}$ and frequency $\omega$.
The first column corresponds to real part and second column corresponds to imaginary part.
The first row corresponds to band gap $\Delta=0.5$ and second row corresponds to $\Delta=1$.
The red arrows indicate the point ${\bf q}=2{\bf k}_{F}$.
}
   \end{center}
\end{figure}
\clearpage
Fig.4
\begin{figure}[!ht]
   \centering
 \centering
   \begin{center}
     \includegraphics*[width=0.7\linewidth]{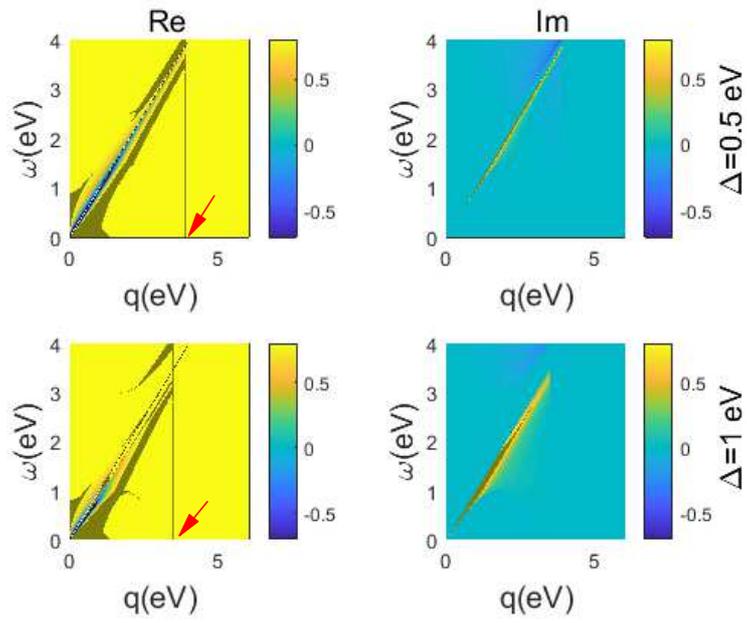}
\caption{(Color online) The dielectric function corresponds to the Fig.3.
The red arrows indicate the point ${\bf q}=2{\bf k}_{F}$.
}
   \end{center}
\end{figure}
\clearpage
Fig.5
\begin{figure}[!ht]
   \centering
 \centering
   \begin{center}
     \includegraphics*[width=0.7\linewidth]{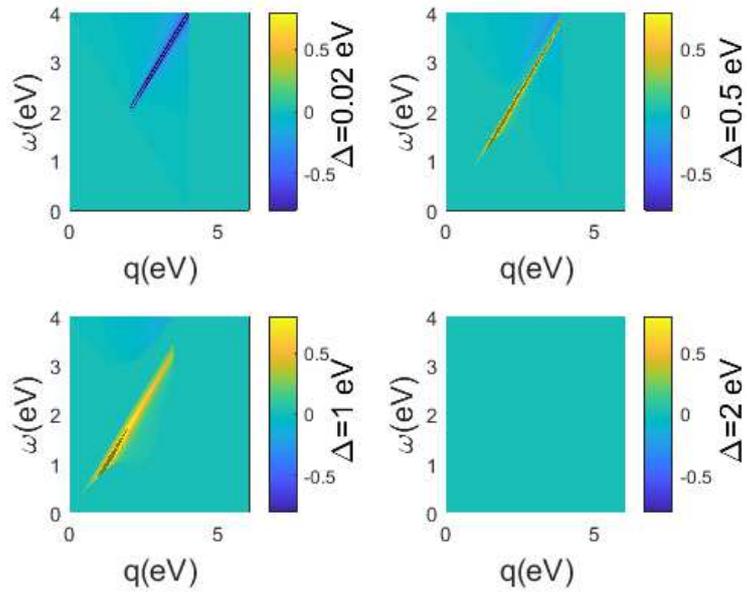}
\caption{(Color online) The absorption of the radiation (not the optical absorbance) as discussed in the text.
The band gaps is corresponds to that in Fig.2.
The red arrows indicate the point ${\bf q}=2{\bf k}_{F}$.
}
   \end{center}
\end{figure}
\clearpage
Fig.6
\begin{figure}[!ht]
   \centering
 \centering
   \begin{center}
     \includegraphics*[width=0.8\linewidth]{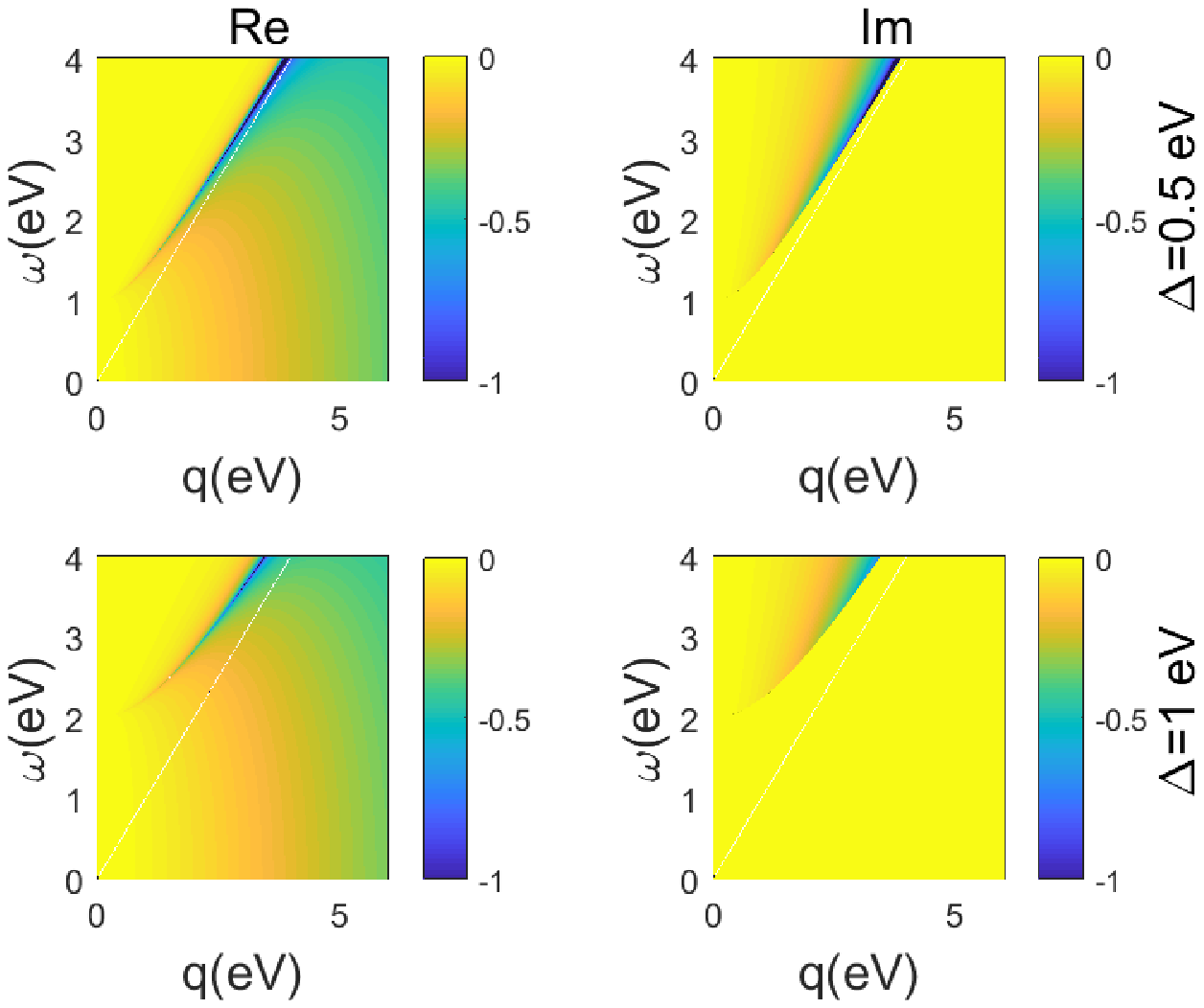}
\caption{(Color online) The same as the Fig.3 but for $\mu=0.01$ eV.
}
   \end{center}
\end{figure}
\clearpage
Fig.7
\begin{figure}[!ht]
   \centering
 \centering
   \begin{center}
     \includegraphics*[width=0.8\linewidth]{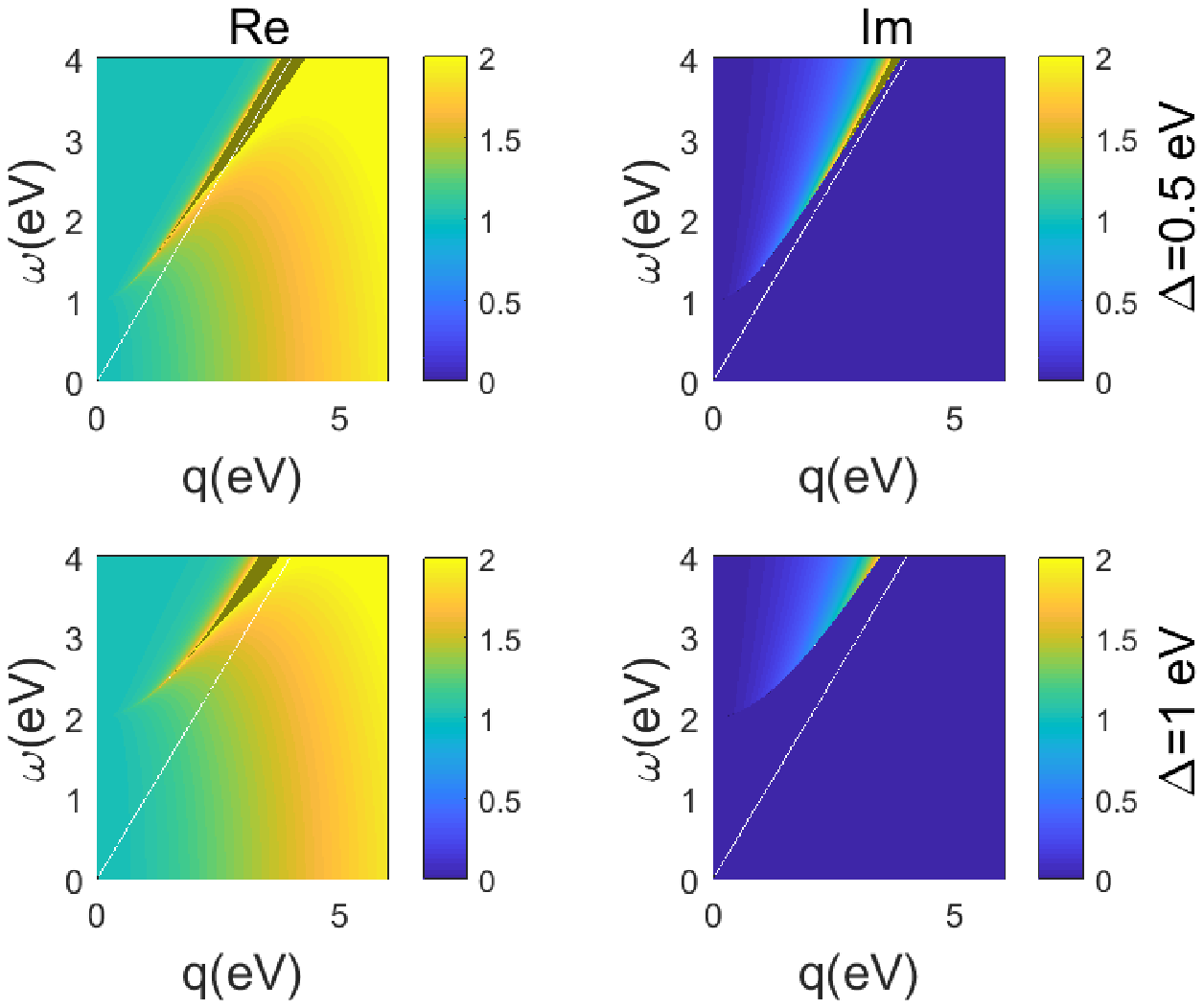}
\caption{(Color online) The same as the Fig.4 but for $\mu=0.01$ eV.
}
   \end{center}
\end{figure}
\clearpage
Fig.8
\begin{figure}[!ht]
   \centering
 \centering
   \begin{center}
     \includegraphics*[width=0.8\linewidth]{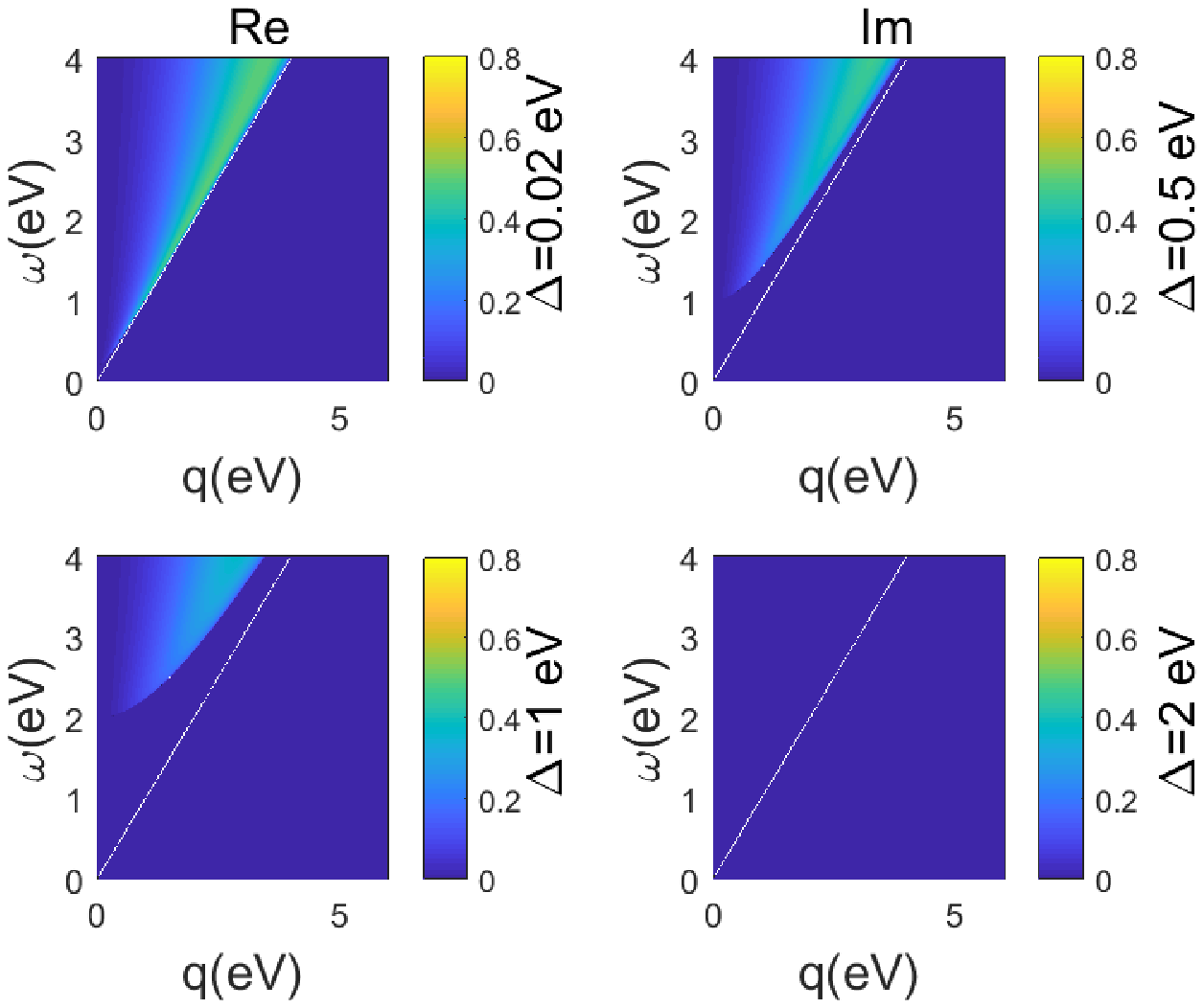}
\caption{(Color online) The same as the Fig.5 but for $\mu=0.01$ eV.
}
   \end{center}
\end{figure}

\clearpage
Fig.9
\begin{figure}[!ht]
   \centering
 \centering
   \begin{center}
     \includegraphics*[width=0.8\linewidth]{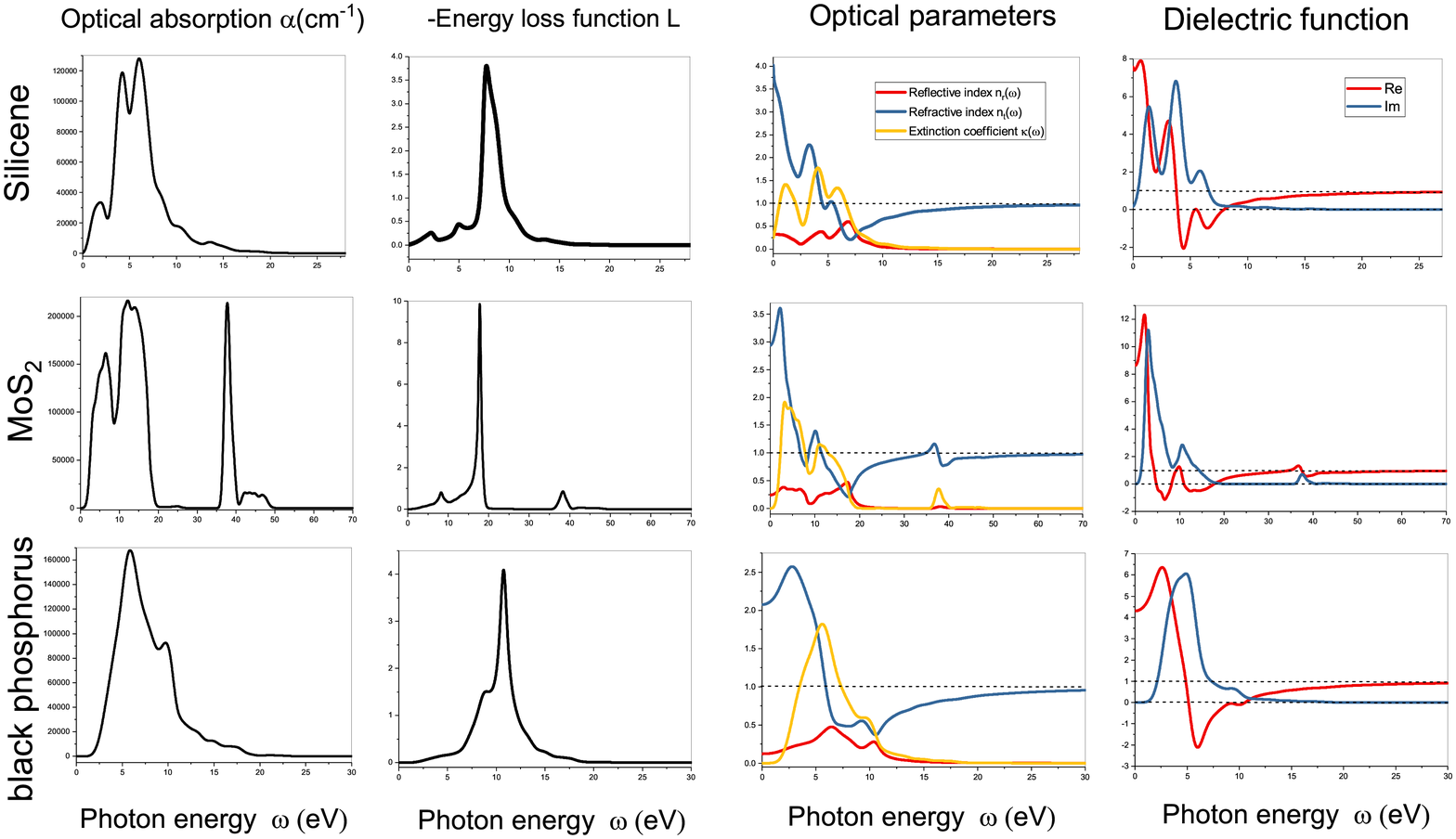}
\caption{(Color online) Optical absorption, inverted energy loss function, optical parameters, and the dielectric function
as a function of the photon energy.
Note that we only show the in-plane component of the above physics quantities.
%{Optical properties of graphene, silicene, germanene, and stanene from IR to far UV –A first principles study}
The first, middel, and last rows are correponds to the monolayer silicene, monolayer MoS$_{2}$,
and monolayer black phosphorus, respectively.
}
   \end{center}
\end{figure}


\begin{thebibliography}{99}
\bibitem{Liu H}Liu H, Neal A T, Zhu Z, et al. Phosphorene: an unexplored 2D semiconductor with a high hole mobility[J]. ACS nano, 2014, 8(4): 4033-4041.
\bibitem{Liu Y}Liu Y, Low T, Ruden P P. Mobility anisotropy in monolayer black phosphorus due to scattering by charged impurities[J]. Physical Review B, 2016, 93(16): 165402.
\bibitem{Wu C HX}Wu C H. Electronic transport and the related anomalous effects in silicene-like hexagonal lattice[J]. arXiv preprint arXiv:1807.10898, 2018.
\bibitem{Wu C H3}Wu C H. Interband and intraband transition, dynamical polarization and screening of the monolayer and bilayer silicene in low-energy tight-binding model[J]. arXiv preprint arXiv:1805.07736, 2018.
\bibitem{Wu C H1}Wu C H. Geometrical structure and the electron transport properties of monolayer and bilayer silicene near the semimetal-insulator transition point in tight-binding model[J]. arXiv preprint arXiv:1805.00350, 2018.
\bibitem{Wu C H4}Wu C H. Integer quantum Hall conductivity and longitudinal conductivity in silicene under the electric field and magnetic field[J]. arXiv preprint arXiv:1805.10656, 2018.
\bibitem{Wu C H7}Wu C H. Josephson effect in silicene-based SNS Josephson junction: Andreev reflection and free energy[J]. arXiv preprint arXiv:1806.10289, 2018.
\bibitem{Wu C H2}Wu C H. Tight-binding model and ab initio calculation of silicene with strong spin-orbit coupling in low-energy limit[J]. arXiv preprint arXiv:1804.01695, 2018.
\bibitem{Wu C H5}Wu C H. Anomalous Rabi oscillation and related dynamical polarizations under the off-resonance circularly polarized light[J]. arXiv preprint arXiv:1806.03592, 2018.

\bibitem{Koutserimpas T T}Koutserimpas T T, Al{\`u} A, Fleury R. Parametric amplification and bidirectional invisibility in PT-symmetric time-Floquet systems[J]. Physical Review A, 2018, 97(1): 013839.
\bibitem{Li C C}Li C C, Xia Q, Lou S Y. Nonlocal Nonlinear Schr?dinger System with Shifted Parity and Delayed Time Reversal Symmetries[J]. Communications in Theoretical Physics, 2018, 70(1): 007.
\bibitem{Perez-Piskunow P M}Perez-Piskunow P M, Torres L E F F, Usaj G. Hierarchy of Floquet gaps and edge states for driven honeycomb lattices[J]. Physical Review A, 2015, 91(4): 043625.
\bibitem{Zhao A}Zhao A, Shen S Q. Quantum anomalous Hall effect in a flat band ferromagnet[J]. Physical Review B, 2012, 85(8): 085209.
\bibitem{Jungwirth T}Jungwirth T, Niu Q, MacDonald A H. Anomalous Hall effect in ferromagnetic semiconductors[J]. Physical review letters, 2002, 88(20): 207208.
\bibitem{Huang R}Huang R, Li J, Wu Z, et al. Universal absorption of two-dimensional materials within k? p method[J]. Physics Letters A, 2018.
\bibitem{Matthes L}Matthes L, Pulci O, Bechstedt F. Optical properties of two-dimensional honeycomb crystals graphene, silicene, germanene, and tinene from first principles[J]. New Journal of Physics, 2014, 16(10): 105007.
\bibitem{Kobayashi K}Kobayashi K, Yabashi M, Takata Y, et al. High resolution-high energy x-ray photoelectron spectroscopy using third-generation synchrotron radiation source, and its application to Si-high k insulator systems[J]. Applied physics letters, 2003, 83(5): 1005-1007.
\bibitem{Matthes L2}Matthes L, Gori P, Pulci O, et al. Universal infrared absorbance of two-dimensional honeycomb group-IV crystals[J]. Physical Review B, 2013, 87(3): 035438.
\bibitem{Kormányos A}Korm{\'a}nyos A, Z{\'o}lyomi V, Drummond N D, et al. Monolayer MoS 2: trigonal warping, the Γ valley, and spin-orbit coupling effects[J]. Physical review b, 2013, 88(4): 045416.

\bibitem{Tikhonenko F V}Tikhonenko F V, Horsell D W, Gorbachev R V, et al. Weak localization in graphene flakes[J]. Physical review letters, 2008, 100(5): 056802.
\bibitem{Feng B}Feng B, Li H, Liu C C, et al. Observation of Dirac cone warping and chirality effects in silicene[J]. ACS nano, 2013, 7(10): 9049-9054.
\bibitem{Fu L}Fu L. Hexagonal warping effects in the surface states of the topological insulator Bi 2 Te 3[J]. Physical review letters, 2009, 103(26): 266801.
\bibitem{John R}John R, Merlin B. Optical properties of graphene, silicene, germanene, and stanene from IR to far UV–a first principles study[J]. Journal of Physics and Chemistry of Solids, 2017, 110: 307-315.
\bibitem{Souslov A}Souslov A, Liu A J, Lubensky T C. Elasticity and response in nearly isostatic periodic lattices[J]. Physical review letters, 2009, 103(20): 205503.
\bibitem{Ezawa M}Ezawa M. Quasi-topological insulator and trigonal warping in gated bilayer silicene[J]. Journal of the Physical Society of Japan, 2012, 81(10): 104713.
\bibitem{McCann E}McCann E, Fal’ko V I. Landau-level degeneracy and quantum Hall effect in a graphite bilayer[J]. Physical Review Letters, 2006, 96(8): 086805.
\bibitem{Morell E S}Morell E S, Torres L E F F. Radiation effects on the electronic properties of bilayer graphene[J]. Physical Review B, 2012, 86(12): 125449.
\bibitem{Pyatkovskiy P K}Pyatkovskiy P K. Dynamical polarization, screening, and plasmons in gapped graphene[J]. Journal of Physics: Condensed Matter, 2008, 21(2): 025506.
\bibitem{Kotov V N}Kotov V N, Uchoa B, Pereira V M, et al. Electron-electron interactions in graphene: Current status and perspectives[J]. Reviews of Modern Physics, 2012, 84(3): 1067.
\bibitem{Wunsch B}Wunsch B, Stauber T, Sols F, et al. Dynamical polarization of graphene at finite doping[J]. New Journal of Physics, 2006, 8(12): 318.
\bibitem{Tabert C J}Tabert C J, Nicol E J. Dynamical polarization function, plasmons, and screening in silicene and other buckled honeycomb lattices[J]. Physical Review B, 2014, 89(19): 195410.
\bibitem{Scholz A}Scholz A, Stauber T, Schliemann J. Dielectric function, screening, and plasmons of graphene in the presence of spin-orbit interactions[J]. Physical Review B, 2012, 86(19): 195424.
\bibitem{Gajdo? M}Gajdo{\v{s}} M, Hummer K, Kresse G, et al. Linear optical properties in the projector-augmented wave methodology[J]. Physical Review B, 2006, 73(4): 045112.
\bibitem{Giannozzi P}Giannozzi P, Baroni S, Bonini N, et al. QUANTUM ESPRESSO: a modular and open-source software project for quantum simulations of materials[J]. Journal of physics: Condensed matter, 2009, 21(39): 395502.
\bibitem{Singh N}Singh N, Kaloni T P, Schwingenschl?gl U. A first-principles investigation of the optical spectra of oxidized graphene[J]. Applied Physics Letters, 2013, 102(2): 023101.
\bibitem{Çakır D}{\c{C}}ak{\i}r D, Sahin H, Peeters F M. Tuning of the electronic and optical properties of single-layer black phosphorus by strain[J]. Physical Review B, 2014, 90(20): 205421.
\bibitem{Tran V}Tran V, Soklaski R, Liang Y, et al. Layer-controlled band gap and anisotropic excitons in few-layer black phosphorus[J]. Physical Review B, 2014, 89(23): 235319.
\bibitem{Wu C H8}Wu C H. Time Evolution and Thermodynamics for the Nonequilibrium System in Phase-Space[J]. arXiv preprint arXiv:1711.00547, 2017.
\bibitem{Yuan S}Yuan S, Rold{\'a}n R, Katsnelson M I. Excitation spectrum and high-energy plasmons in single-layer and multilayer graphene[J]. Physical Review B, 2011, 84(3): 035439.
%\bibitem{Yang K}Yang K, Cahangirov S, Cantarero A, et al. Thermoelectric properties of atomically thin silicene and germanene nanostructures[J]. Physical Review B, 2014, 89(12): 125403.
%\bibitem{Gürel H H}G{\"u}rel H H, {\"O}z{\c{c}}elik V O, Ciraci S. Effects of charging and perpendicular electric field on the properties of silicene and germanene[J]. Journal of Physics: Condensed Matter, 2013, 25(30): 305007.
\bibitem{Yin X}Yin X, Ye Z, Chenet D A, et al. Edge nonlinear optics on a MoS2 atomic monolayer[J]. Science, 2014, 344(6183): 488-490.
\bibitem{Yang L}Yang L, Deslippe J, Park C H, et al. Excitonic effects on the optical response of graphene and bilayer graphene[J]. Physical review letters, 2009, 103(18): 186802.

\bibitem{González J}Gonz{\'a}lez J, Guinea F, Vozmediano M A H. Marginal-Fermi-liquid behavior from two-dimensional Coulomb interaction[J]. Physical Review B, 1999, 59(4): R2474.
\bibitem{Khveshchenko D V}Khveshchenko D V. Ghost excitonic insulator transition in layered graphite[J]. Physical Review Letters, 2001, 87(24): 246802.



%\bibitem{Guzmn-Verri G G}Guzmn-Verri G G, Voon L C L Y. Electronic structure of silicon-based nanostructures. Physical Review B, 2007, 76(7): 075131.


\end{thebibliography}
\end{document}